\documentclass[letter]{aa}
\usepackage{txfonts}
\usepackage{graphicx}
\begin{document}

\title{The environment of the fast rotating star Achernar}
\subtitle{High-resolution thermal infrared imaging with VISIR in BURST mode}
\titlerunning{Thermal infrared imaging of Achernar}
\authorrunning{P. Kervella \& A. Domiciano de Souza}
\author{
P. Kervella\inst{1}
\and
A. Domiciano~de~Souza\inst{2,3}
}
\offprints{P. Kervella}
\mail{Pierre.Kervella@obspm.fr}
\institute{
LESIA, Observatoire de Paris, CNRS, UPMC, Universit\'e Paris Diderot, 5 Place Jules Janssen,
F-92195 Meudon Cedex, France
\and
Lab. Univ. d'Astrophysique de Nice (LUAN), CNRS UMR 6525, UNSA, Parc Valrose,
F-06108 Nice, France
\and
Observatoire de la C\^ote d'Azur, CNRS UMR 6203, D{\'e}partement GEMINI,
BP 4229, F-06304 Nice Cedex 4, France
}
\date{Received ; Accepted}
\abstract
{It is expected that Be stars are surrounded by circumstellar envelopes, and that a significant fraction have companions. Achernar ($\alpha$\,Eri) is the nearest Be star, and is thus a favourable target to search for their signatures using high resolution imaging.}
{We aim at detecting circumstellar material or companions around Achernar at distances of a few tens of AU.}
{We obtained diffraction-limited thermal IR images of Achernar using the BURST mode of the VLT/VISIR instrument.}
{The images obtained in the PAH1 band show a point-like secondary source located 0.280" north-west of Achernar. Its emission is 1.8\% of the flux of Achernar in this band, but is not detected in the PAH2, SiC and NeII bands. }
{The flux from the detected secondary source is compatible with a late A spectral type main sequence companion to Achernar. The position angle of this source (almost aligned with the equatorial plane of Achernar) and its projected linear separation (12.3\,AU at the distance of Achernar) favor this interpretation.}

\maketitle

\section{Introduction}

The southern star \object{Achernar} ($\alpha$\,Eridani, \object{HD 10144}) is the brightest and nearest of all Be stars (V $=0.46$ mag). Depending on the author (and the technique used) its spectral type ranges from B3-B4IIIe to B4Ve (e.g., Slettebak~\cite{slettebak82}, Balona et al.~\cite{balona87}). The estimated projected rotation velocity $v \sin i$ ranges from 220 to 270\,km.s$^{-1}$ and the effective temperature $T_{\rm eff}$ from $15\,000$ to $20\,000$\,K (see e.g., Vinicius et al.~\cite{vinicius06}, Rivinius \textit{priv. comm.}, Chauville et al.~\cite{chauville01}).
It has been the subject of a renewed interest since its distorted photosphere was resolved by means of near-IR long-baseline interferometry (Domicano de Souza et al.~\cite{domiciano03}). Further interferometric observations revealed the polar wind ejected from the hot polar caps of the star (Kervella \& Domiciano~\cite{kervella06}), due to the von Zeipel effect (von Zeipel~\cite{vonzeipel24}).

Our observations aim at studying two aspects of the close environment of Achernar: the circumstellar envelope at distances of up to a few tens of AU, and binarity. Mid-infrared (hereafter MIR) imaging is prefectly suited for these two objectives. Firstly, this wavelength range corresponds to where a circumstellar envelope becomes optically thick (asuming the emission is caused by free-free radiation, following for instance Panagia \& Felli~\cite{panagia75}). Secondly, the contrast between Achernar and a cool companion will be significantly reduced.

We hereafter present the result of our imaging campaign with the VLT/VISIR instrument, to explore the environment of this star at angular distances of $\approx 0.1$ to 10".

\section{Observations \label{observations}}

\subsection{Instrumental setup: the BURST mode of VISIR \label{instrument}}

We used the VISIR instrument (Lagage et al.~\cite{lagage04}), installed at the Cassegrain focus of the Melipal telescope (UT3) of the VLT (Paranal, Chile). VISIR is a MIR imager, that also provides a slit spectrometer. As it is a ground-based instrument, its sensitivity is severely limited by the high thermal background of the atmosphere, compared for instance to the Spitzer space telescope, but its resolving power is ten times higher, thanks to the 8\,m diameter of the primary mirror. VISIR is therefore very well suited for our programme to search for the presence of circumstellar material and companions within a few tens of AU of Achernar.

However, under standard conditions at Paranal (median seeing of 0.8" at 0.5\,$\mu$m), the 8\,m telescope is not diffraction limited in the MIR (seeing $\approx 0.4"$ vs. 0.3" diffraction). Instead of a pure Airy diffraction pattern, several moving speckles and tip-tilt usually degrade the quality of the image (see e.g. Tokovinin, Sarazin \& Smette~\cite{tokovinin07}). To overcome this limitation, a specific mode of the instrument, called the BURST mode, was introduced by Doucet et al.~(\cite{doucet07a}, \cite{doucet07b}). Its principle is to acquire very short exposures ($\Delta t \lesssim 50$\,ms), in order to keep the complete integration within a fraction of the coherence time ($\approx 300$\,ms at Paranal in the MIR).
The detector is therefore read very quickly, and the resulting images freeze the turbulence. It is subsequently possible to select the best images that present a single speckle (``lucky imaging''), and are thus diffraction-limited. The details of this selection procedure are given in Sect.~\ref{datared}. The BURST mode was already applied successfully to the observation of the nucleus of NGC\,1068 (Poncelet et al.~\cite{poncelet07}).

\subsection{Observations log \label{log}}

We observed Achernar during the first half of the night of October 3-4, 2006. A series of BURST mode observations of this star and $\delta$\,Phe was obtained in three filters: PAH1, PAH2 and NeII (central wavelengths $\lambda = 8.59$, 11.25 and 12.81\,$\mu$m). They were followed by classical VISIR imaging observations of Achernar and $\delta$\,Phe in the PAH1, SiC ($\lambda = 11.85\,\mu$m) and NeII filters, in order to provide a reference for processing using the standard instrument pipeline.

The detailed transmission curves of the filters can be found in the VISIR instrument manual, available from the ESO web site\footnote{http://www.eso.org/instruments/visir}. The images obtained on the targets were chopped and nodded in a North-South and East-West direction by offsetting the M2 mirror and the telescope itself (resp.) in order to remove the fluctuating thermal background in the post-processing. The chopping and nodding amplitudes were both set to 8" on the sky, and the periods were 4 and 90\,s, respectively. The pixel scale was set to the smallest scale available on VISIR (0.075"/pixel), in order to sample as well as possible the $\approx$0.3" FWHM diffraction pattern of the telescope.
The Achernar observations were interspersed every 40-50 minutes with a PSF reference star, observed with the same instrumental setup. The journal of the VISIR BURST mode observations is given in Table~\ref{visir_log}. During the observations, the seeing quality in the visible varied from average (1.1") to excellent (0.6").

\begin{table}
\caption{Log of the observations of Achernar and its PSF calibrator, $\delta$\,Phe using the BURST mode of VISIR. MJD is the modified Julian date of the middle of the exposures on the target, minus 54\,012. The Detector Integration Time (DIT) is given in milliseconds for one BURST image. $\theta$ is the seeing in the visible ($\lambda=0.5\,\mu$m) as measured by the observatory DIMM sensor, in arcseconds. The airmass (AM) is the average airmass of the observation.} 
\label{visir_log}
\begin{small}
\begin{tabular}{cccccccc}
\hline
\# & MJD$^*$ & Star & Filter & DIT & $N$ exp. & $\theta$\,(") & AM \\
\hline
A & 0.0669 & $\alpha$\,Eri & PAH1 & 20 & 1200 $\times$ 14 & 0.9 & 1.64 \\
B & 0.0844 & $\delta$\,Phe & PAH1 & 20 & 1200 $\times$ 14 & 1.0 & 1.43 \\
C & 0.0965 & $\delta$\,Phe & NeII & 25 & 960 $\times$ 14 & 1.1 & 1.37 \\
D & 0.1201 & $\alpha$\,Eri & NeII & 25 & 960 $\times$ 14 & 1.0 & 1.36 \\
E & 0.1319 & $\alpha$\,Eri & PAH1 & 20 & 1200 $\times$ 14 & 0.6 & 1.32 \\
F & 0.1482 & $\delta$\,Phe & PAH1 & 20 & 1200 $\times$ 14 & 0.6 & 1.18 \\
G & 0.1604 & $\delta$\,Phe & PAH2 & 20 & 1200 $\times$ 14 & 0.6 & 1.16 \\
H & 0.1877 & $\alpha$\,Eri & PAH2 & 20 & 1200 $\times$ 14 & 0.8 & 1.21 \\
I & 0.1997 & $\alpha$\,Eri & PAH1 & 20 & 1200 $\times$ 14 & 0.8 & 1.20 \\
\hline
\end{tabular}
\end{small}
\end{table}

\begin{table}
\caption{Log of the observations of Achernar and $\delta$\,Phe using the classical imaging mode of VISIR. The columns are the same as in Table~\ref{visir_log}, except the ``Total" column, giving the total integration time on target.} 
\label{visir_log_service}
\begin{small}
\begin{tabular}{cccccccc}
\hline
\# & MJD$^*$ & Star & Filter & DIT & Total\,(s) & $\theta$\,(") & AM \\
\hline
J & 0.2085 & $\alpha$\,Eri & SiC & 25 & 640 & 0.7 & 1.19 \\
K & 0.2188 & $\alpha$\,Eri & NeII & 25 & 650 & 0.7 & 1.19 \\
L & 0.2292 & $\alpha$\,Eri & PAH1 & 20 & 650 & 0.7 & 1.19 \\
M & 0.2445 & $\delta$\,Phe & SiC & 25 & 640 & 0.7 & 1.10 \\
N & 0.2474 & $\delta$\,Phe & NeII & 25 & 650 & 0.6 & 1.11 \\
O & 0.2518 & $\delta$\,Phe & PAH1 & 20 & 650 & 0.6 & 1.11 \\
\hline
\end{tabular}
\end{small}
\end{table}

The PSF reference star, $\delta$\,Phe (\object{HD 9362}, G9III), was chosen in the Cohen et al.~(\cite{cohen99}) catalogue of spectrophotometric standards for infrared wavelengths. In addition to being a stable star, previous interferometric observations in the near-IR with the VINCI instrument (see in particular Th\'evenin et al.~\cite{thevenin05} and Kervella \& Domiciano de Souza~\cite{kervella06}) have confirmed that $\delta$\,Phe is indeed single. Another advantage of choosing our PSF star in this catalogue is that its flux is absolutely calibrated, and it therefore provides a convenient photometric reference to estimate accurately the absolute flux of the observed object. $\delta$\,Phe is located relatively close to Achernar on the sky ($8.2^\circ$), and therefore at a similar airmass. It is also of comparable brightness in the MIR (9.5\,Jy at 12$\mu$m vs. $\approx 16$\,Jy for Achernar).

\subsection{Raw data processing \label{datared}}

One drawback of the BURST mode is that it produces a very large quantity of data: almost 20\,Gbytes for our half-night of observations. However, the basic MIR data reduction to remove the instrument signature and the thermal background from the images is simple, and the data quantity is quickly brought to a more manageable volume. The fluctuations of the thermal background were removed through the classical subtraction of the chopped and nodded images, in order to produce data cubes of more than 10\,000 images covering $6.8 \times 6.8$".
After a precentering at the integer pixel level, the images were sorted based on their maximum intensity, used as a proxy of the Strehl ratio. The 5\,000 best images of each cube were then resampled up by a factor 10 using a cubic spline interpolation, and the star image was subsequently centered using gaussian fitting, at a precision level of a few milliarcseconds. The field of view was trimmed to $2.25 \times 2.25$" to reduce the computing time. The resulting cubes were eventually averaged to obtain the master images of Achernar and $\delta$\,Phe used in the image analysis process described below.

Our classical (non-BURST) VISIR images were processed using the standard VISIR pipeline. This mode differs from the BURST mode in the sense that the images obtained within one chopping cycle (every 90\,seconds) are averaged together immediately after acquisition in order to reduce the data rate. As a result, the atmospheric tip-tilt and higher order perturbations degrade the image quality. The main advantage of this mode is that the processing is less computer intensive.

\subsection{BURST vs. classical mode images}

\begin{figure}[t]
\centering
\includegraphics[width=8.9cm]{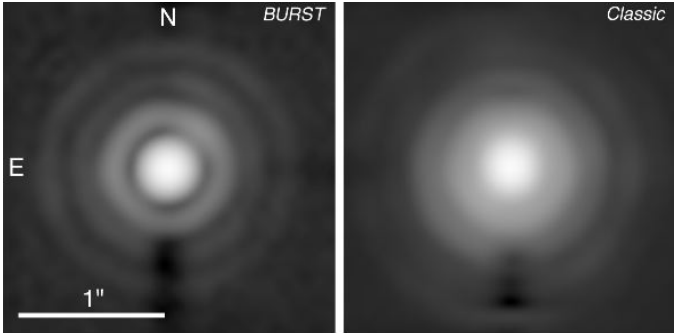}
\caption{Comparison of the BURST (image A in Table~\ref{visir_log}) and Classic (image L in Table~\ref{visir_log_service}) images of Achernar in the PAH1 filter. The two images were obtained in the same seeing conditions.}
\label{Burst-Classic}
\end{figure}

As expected, the classical mode images present a significantly degraded effective resolution compared to the BURST mode images (Fig.~\ref{Burst-Classic}), although the seeing conditions were excellent in both cases (visible seeing $\simeq 0.7$"). The full-width at half-maximum (hereafter FWHM) of the classical mode image is 0.285", while it is only 0.217" for the BURST image. As already demonstrated by Doucet et al.~(\cite{doucet07a}), this clearly shows that the BURST mode is perfectly suited for the most demanding observing programs in terms of angular resolution.

\section{Image analysis}

\subsection{PSF subtraction and image deconvolution}

To analyse the images of Achernar, we used two distinct methods: PSF subtraction and Lucy-Richardson deconvolution. Provided the seeing conditions are comparable between Achernar and the calibrator (a condition verified during our observations), these two methods give access to the close environment of the star. The images were processed pairwise with the following Achernar-PSF combinations (see Table~\ref{visir_log}): {\bf A-B} (PAH1), {\bf D-C} (NeII), {\bf E-F} (PAH1), {\bf H-G} (PAH2), {\bf I-F} (PAH1). 

\begin{figure*}[tbp]
\centering
\includegraphics[width=8.0cm]{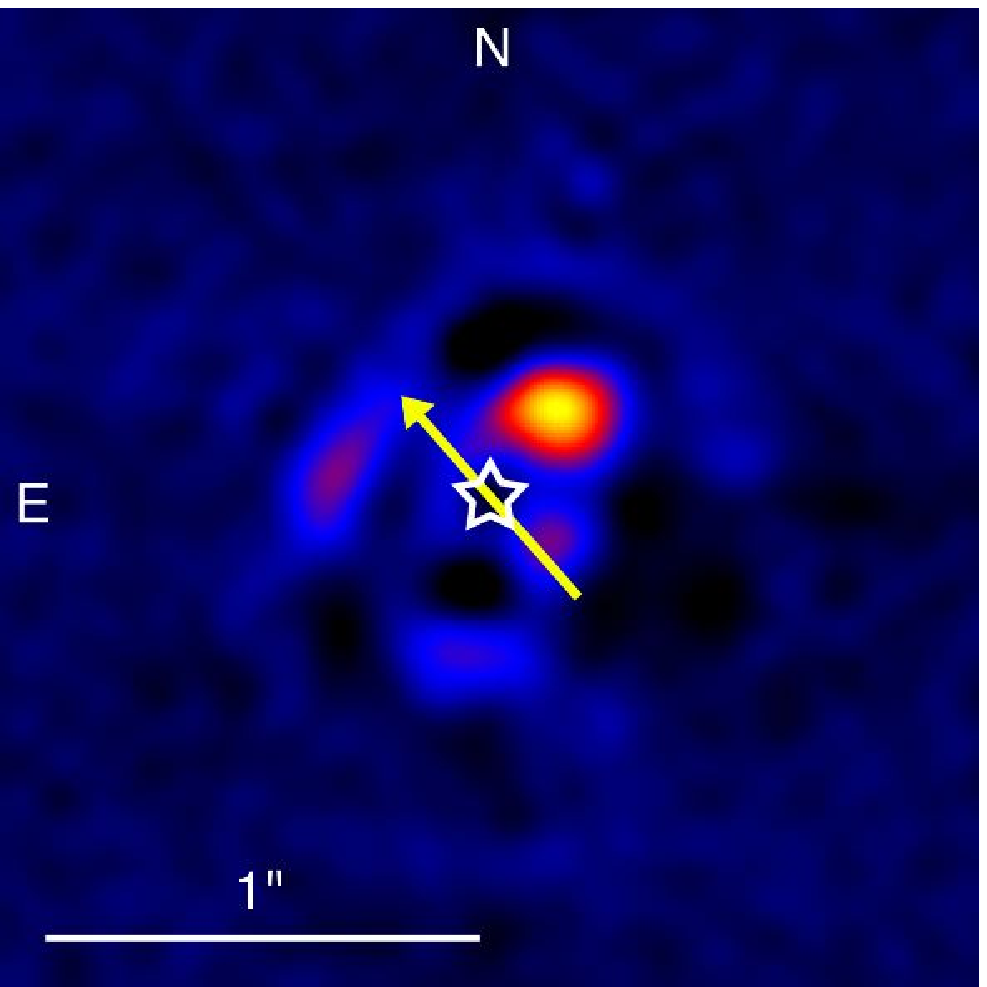} \hspace{1.2cm}
\includegraphics[width=8.0cm]{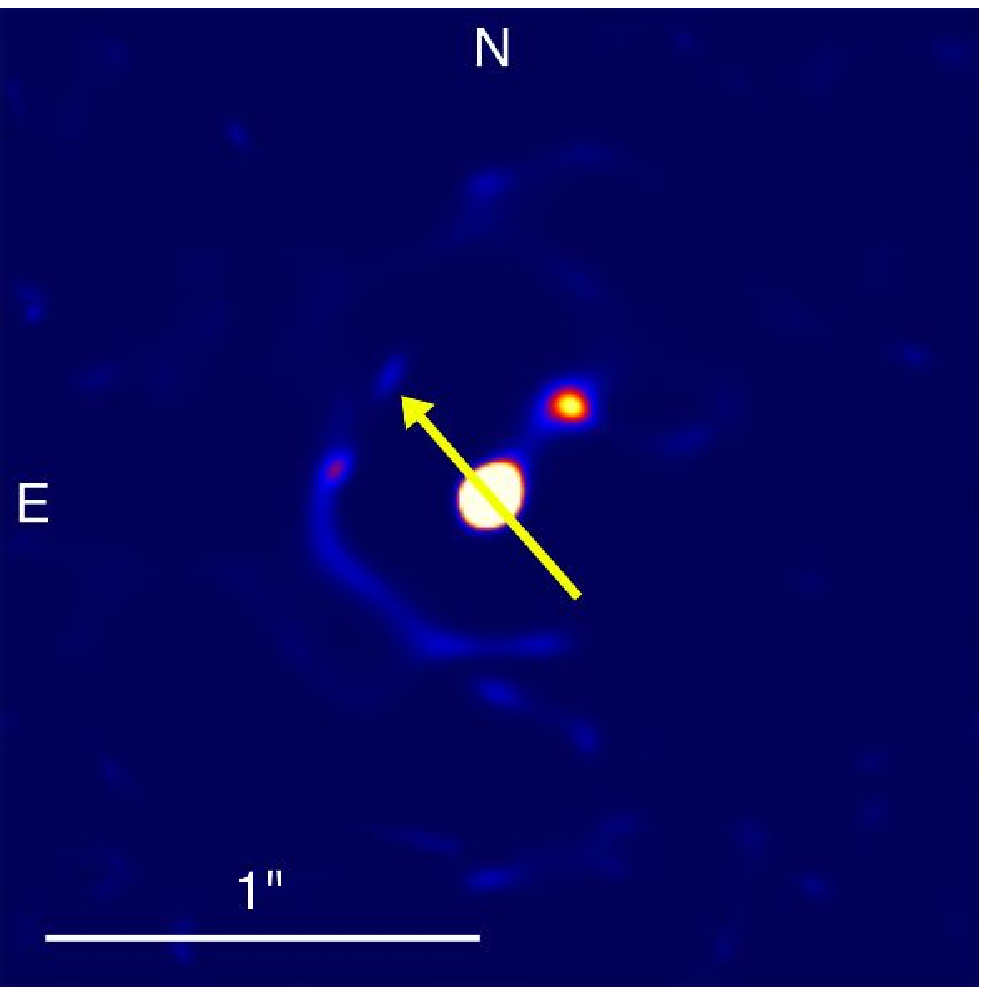}
\caption{{\it Left:} Average of the three PSF subtracted images of Achernar in the PAH1 filter (AB, EF and IF pairs). The position of Achernar is marked with a ``star" symbol. {\it Right:} Average deconvolved image of Achernar (without PSF subtraction). In both images, the orientation of the polar axis of Achernar (Kervella \& Domiciano~\cite{kervella06}) is represented by an arrow.}
\label{AvgPAH1}
\end{figure*}

\begin{figure}[tbp]
\centering
\includegraphics[width=8.9cm]{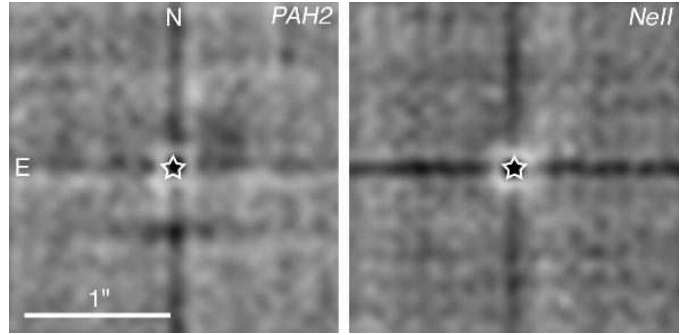}
\caption{Residual of the subtraction of the PSF calibrator image from the images of Achernar obtained using the PAH2 (D-C) and NeII (H-G) filters ($\lambda =$ 11.25 and 12.81\,$\mu$m, respectively).
The vertical and horizontal dark lines are artefacts from the detector.}
\label{NeII-PAH2}
\end{figure}

Firstly, we directly subtracted the normalized PSF image obtained on the calibrator star from the image of Achernar. The normalization factor between the two images was computed using a two-parameter linear least squares fit (slope and zero point) on the central $375\times375$\,mas ($50\times50$\,pixels on our resampled images).
A higher degree fit was also tried to account for a possible non-linearity of the detector, but no difference was noticeable between the two fitting methods.
The average of the three subtracted PAH1 images is presented in Fig.~\ref{AvgPAH1} (left), and the NeII and PAH2 images are shown in Fig.~\ref{NeII-PAH2}.
This procedure does not work well with the classical (non-BURST) images, due to the degraded image quality. We thus employed the classical mode images only for photometry.

Secondly, we deconvolved the three Achernar images using their associated PSF calibrator images as the ``dirty beam" and the classical Lucy deconvolution algorithm (200 iterations). The average of the three resulting deconvolved images in the PAH1 band is presented in Fig.~\ref{AvgPAH1} (right). The deconvolved images in the NeII and PAH2 bands do not show any source in the field around Achernar.


\subsection{Photometry of Achernar \label{photom}}

We obtained photometry of Achernar and $\delta$\,Phe using an aperture of 0.6" in diameter. In order to absolutely calibrate the flux of Achernar in the four filters used for the present observations, we used as reference the spectrophotometric template of $\delta$\,Phe from Cohen et al.~(\cite{cohen99}). The irradiance of $\delta$\,Phe was read from this template at the average wavelength of each filter, and multiplied by the ratio of the measured aperture photometry of Achernar and $\delta$\,Phe. The results are presented in Table~\ref{photom_achernar}.

\begin{table}
\caption{Measured irradiance of Achernar.}
\label{photom_achernar}
\begin{tabular}{lccc}
\hline
Filter & $\lambda$ ($\mu$m) & $10^{-13}$ W/m$^2$/$\mu$m & Jy \\
\hline
\noalign{\smallskip}
PAH1 & 8.59 & 9.99 & 23.7\\
PAH2 & 11.25 & 4.01 & 16.8\\
SiC & 11.85 & 3.06 & 14.4\\
NeII & 12.81 & 2.31 & 12.5\\
\hline
\end{tabular}
\end{table}

\subsection{Northwestern emission}

An emitting region in the northwest quadrant, hereafter refered to as ``source B", is visible in the subtracted and deconvolved PAH1 images (Fig.~ \ref{AvgPAH1}).
A gaussian fit to the deconvolved image of Achernar gives the following relative position of source B from Achernar along the right ascension and declination directions:
$\Delta \alpha = -0.184",\ \Delta \delta = +0.211"$,
giving an angular separation of 0.280" and a position angle $\phi = -41.1^\circ$ (counted negatively from North towards West, for which $\phi = -90^\circ$). At the distance of Achernar (44\,pc, ESA~\cite{hip}) this corresponds to a linear projected separation of 12.3\,AU. The measured azimuth is close to that of the equatorial plane of Achernar: $\phi_{\rm eq} = -48.4^\circ$ (Kervella \& Domiciano de Souza~\cite{kervella06}).

The FWHM of source B on the subtracted image (Fig.~\ref{AvgPAH1}) is
$\sigma (\alpha) = 0.243",\ \sigma (\delta) = 0.170"$.
This extension is comparable to the FWHM of the image of Achernar (0.217"), and therefore compatible with a point-like source. This is confirmed by the point-like appearance of the source in the deconvolved image (Fig.~\ref{AvgPAH1}). The peak intensity of source B on the subtracted image reaches 1.95\% of that of Achernar, and an integration over a 0.22" diameter circular aperture gives a flux ratio of 1.79\%. Considering the photometry of Achernar presented in Sect.~\ref{photom}, the absolute flux from source B is therefore $\simeq 1.9\ 10^{-14}$\,W/m$^2$/$\mu$m, or 0.4\,Jy, in the PAH1 band.

\section{Discussion \label{discussion}}

\begin{figure}[tbp]
\centering
\includegraphics[bb=0 0 360 180, width=8.7cm]{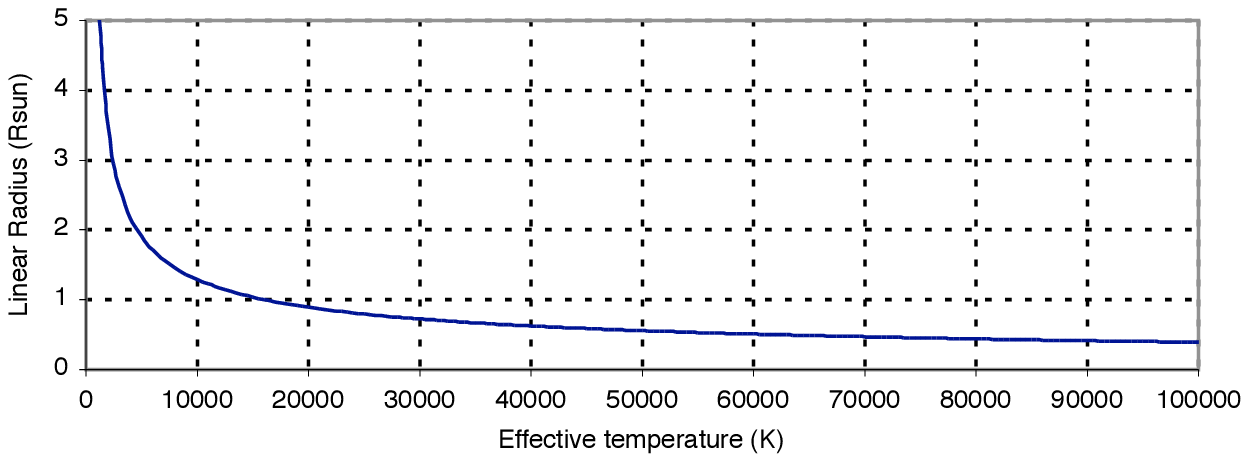}
\caption{Linear radius of the companion of Achernar, as a function of its effective temperature.}
\label{Simu_Companion}
\end{figure}

The nature of source B cannot be established as we have only PAH1 photometry, but Fig~\ref{Simu_Companion} gives the linear radius of the stars that could produce the flux observed in the PAH1 band, as a function of their effective temperature. An interesting possibility is that source B could be an evolved helium star (e.g. a planetary nebula nucleus variable) with $T_{\rm eff} \simeq 10^5$\,K and $R \simeq 0.4\,R_\odot$. Future imaging and spectroscopy will provide a more secure identification, but for the present discussion, we make the hypothesis that it is a main sequence (MS) star. 
At the distance of Achernar, a MS star with an apparent flux of 0.4\,Jy ($m_N \simeq 5.2$, $M_N  \simeq 2.0$) has an effective temperature of $\approx$7\,500\,K (spectral type $\approx$A7V), and a mass of $\approx 2\,M_{\odot}$. Its extrapolated magnitude in the $V$ band is $m_V \simeq 5.8$, giving a contrast of $\Delta m_V = 5.4$ magnitudes with Achernar. The apparent position of this source, almost orthogonal to the polar axis of Achernar, is a clue that its orbit is probably coplanar with the equatorial plane of the star. In turn, this indicates that the polar axis of Achernar could be almost in the plane of the sky.
The non-detection of source B in the PAH2 and NeII filters can be explained by the higher noise in these images and the decreasing flux of the source with increasing wavelength. The residual noise in the PSF subtracted images in the PAH1, PAH2 and NeII bands correspond to 5$\sigma$ point-source detection limits of 0.05, 0.31 and 0.59\,Jy, respectively, while the expected fluxes of an A7V star are 0.42, 0.25 and 0.20\,Jy in the same filters. A 5\,$\sigma$ detection in the PAH2 and NeII bands thus appears difficult.

Statistically, the presence of a companion around a massive star like Achernar is not unexpected: about two-thirds of the stars in this mass range ($M_{\alpha\, \rm Eri} \simeq 6 M_{\odot}$) have at least one companion (Preibisch et al.~\cite{preibisch01}, see also Vanbeveren et al.~\cite{vanbeveren98}). It is also interesting to remark that the estimated binary frequency for Cepheids, a state into which Achernar will evolve in the future, is also around two-third (Szabados~\cite{szabados03}). The detected companions to these supergiants are typically A-type dwarfs, a scenario compatible with the secondary source we detected.

The spectral type and rotational velocity of $\alpha$\,Ara (B3Ve, $v.\sin i \approx 250-300$\,km.s$^{-1}$; Yudin~\cite{yudin01}; Chauville et al.~\cite{chauville01}) make it an interesting analogue of Achernar, though with a higher absolute flux in the MIR. Interferometric observations with the VLTI/MIDI (Chesneau et al.~\cite{chesneau05}) MIR instrument showed that $\alpha$\,Ara is surrounded by a truncated circumstellar disk. This points at the presence of a stellar companion orbiting at a distance of $\approx 30\,R_*$. The detected source B is located much further away, but the presence of companions in both cases could be an indication that the Be phenomenon is linked to binarity.

\section{Conclusion}

We presented high resolution thermal infrared images of the close environment of Achernar. A point-like source is identified at an angular distance of 0.280" from Achernar, contributing for 1.8\% of the flux of Achernar in the PAH1 band. This source is not detected in our longer wavelength images, probably due to their insufficient sensitivity. Its location, close to Achernar and almost in the equatorial plane of the star, indicates that it is likely to be a physical companion orbiting Achernar. The measured flux corresponds to that of an A7V star (similar to Altair, for instance). A secure identification requires further observations, that will also determine if source B indeed shares the proper motion of Achernar ($\simeq 97$\,mas/yr).

\begin{acknowledgements}
We wish to thank the VISIR instrument team at Paranal, and in particular Dr. L. Vanzi, for making the BURST mode available to us in visitor mode, on a very short notice. Based on observations made with ESO Telescopes at Paranal Observatory under programs 078.D-0295(A) and (B). This research used the SIMBAD and VIZIER databases at the CDS, Strasbourg (France), and NASA's ADS bibliographic services.
\end{acknowledgements}
{}

\end{document}